\documentclass[twocolumn,english,aps,pra,10pt,superscriptaddress,floatfix]{revtex4-2}
\usepackage{array}[=2016-10-06]
\usepackage{xcolor}

\usepackage{amsmath}
\usepackage{amsfonts}
\usepackage{stmaryrd}
\usepackage{amssymb}
\usepackage{graphicx}
\usepackage{dcolumn}
\usepackage{multirow}
\usepackage{bm}
\usepackage{changes}
\usepackage{chemformula}
\usepackage[version=4]{mhchem}
\usepackage{soul}
\usepackage{float}
\definecolor{fancyblue}{RGB}{70,120,200}
\definecolor{fancymagenta}{RGB}{200,50,130}
\usepackage[colorlinks=true,citecolor=fancyblue,linkcolor=fancymagenta]{hyperref}

\usepackage[english]{babel}
\usepackage{url}
\usepackage{siunitx}
\DeclareSIUnit \dbc {dBc}
\ExplSyntaxOn
\keys_if_exist:nnTF { siunitx } { reset-text-series }
  { \sisetup{reset-text-series=false, text-series-to-math=true, reset-text-family=false, text-family-to-math=true} }
  { \sisetup{detect-weight=true, detect-family=true} }
\ExplSyntaxOff
\usepackage{microtype}

\begingroup
\makeatletter
\def\XR@@dURL#1#2#3#4#5\\{\unexpanded{{#1}{#2}{#3}{#4}}{\XR@URL}}
\makeatother
\endgroup

\usepackage{times}
\usepackage{titlesec}
\titlespacing*{\subsection}{0pt}{0.8em}{0.3em}

\newenvironment{manuscriptsummary}{\par\noindent\bfseries}{\par\medskip}

\begin{document}
\title{Neodymium-implanted photonic integrated amplifiers, tunable lasers and femtosecond oscillators}
\author{Jiale Sun}
\thanks{These authors contributed equally.}
\affiliation{Institute of Physics, Swiss Federal Institute of Technology Lausanne (EPFL), CH-1015 Lausanne, Switzerland}
\affiliation{Institute of Electrical and Micro Engineering, Swiss Federal Institute of Technology Lausanne (EPFL), CH-1015 Lausanne, Switzerland}
\author{Xinru Ji}
\thanks{These authors contributed equally.}
\affiliation{Institute of Physics, Swiss Federal Institute of Technology Lausanne (EPFL), CH-1015 Lausanne, Switzerland}
\affiliation{Institute of Electrical and Micro Engineering, Swiss Federal Institute of Technology Lausanne (EPFL), CH-1015 Lausanne, Switzerland}
\author{Sören Lerner}
\affiliation{Institute of Solid State Physics, Friedrich Schiller University Jena, 07743 Jena, Germany}
\author{Zheru Qiu}
\affiliation{Institute of Physics, Swiss Federal Institute of Technology Lausanne (EPFL), CH-1015 Lausanne, Switzerland}
\affiliation{Institute of Electrical and Micro Engineering, Swiss Federal Institute of Technology Lausanne (EPFL), CH-1015 Lausanne, Switzerland}
\author{Yanwu Liu}
\affiliation{Institute of Physics, Swiss Federal Institute of Technology Lausanne (EPFL), CH-1015 Lausanne, Switzerland}
\affiliation{Institute of Electrical and Micro Engineering, Swiss Federal Institute of Technology Lausanne (EPFL), CH-1015 Lausanne, Switzerland}
\author{Xuan Yang}
\affiliation{Institute of Physics, Swiss Federal Institute of Technology Lausanne (EPFL), CH-1015 Lausanne, Switzerland}
\affiliation{Institute of Electrical and Micro Engineering, Swiss Federal Institute of Technology Lausanne (EPFL), CH-1015 Lausanne, Switzerland}
\author{Yuxuan Xie}
\affiliation{Institute of Physics, Swiss Federal Institute of Technology Lausanne (EPFL), CH-1015 Lausanne, Switzerland}
\affiliation{Institute of Electrical and Micro Engineering, Swiss Federal Institute of Technology Lausanne (EPFL), CH-1015 Lausanne, Switzerland}
\author{Annkathrin Köhler}
\affiliation{Institute of Solid State Physics, Friedrich Schiller University Jena, 07743 Jena, Germany}
\author{Lucas Kyncl}
\affiliation{Institute of Solid State Physics, Friedrich Schiller University Jena, 07743 Jena, Germany}
\author{Marcus Albrechtsen}
\affiliation{Institute of Physics, Swiss Federal Institute of Technology Lausanne (EPFL), CH-1015 Lausanne, Switzerland}
\affiliation{Institute of Electrical and Micro Engineering, Swiss Federal Institute of Technology Lausanne (EPFL), CH-1015 Lausanne, Switzerland}
\author{Xurong Li}
\affiliation{Institute of Physics, Swiss Federal Institute of Technology Lausanne (EPFL), CH-1015 Lausanne, Switzerland}
\affiliation{Institute of Electrical and Micro Engineering, Swiss Federal Institute of Technology Lausanne (EPFL), CH-1015 Lausanne, Switzerland}
\author{Yichi Zhang}
\affiliation{Institute of Physics, Swiss Federal Institute of Technology Lausanne (EPFL), CH-1015 Lausanne, Switzerland}
\affiliation{Institute of Electrical and Micro Engineering, Swiss Federal Institute of Technology Lausanne (EPFL), CH-1015 Lausanne, Switzerland}
\author{Carsten Ronning}
\affiliation{Institute of Solid State Physics, Friedrich Schiller University Jena, 07743 Jena, Germany}
\author{Tobias J. Kippenberg}
\email{tobias.kippenberg@epfl.ch}
\affiliation{Institute of Physics, Swiss Federal Institute of Technology Lausanne (EPFL), CH-1015 Lausanne, Switzerland}
\affiliation{Institute of Electrical and Micro Engineering, Swiss Federal Institute of Technology Lausanne (EPFL), CH-1015 Lausanne, Switzerland}
\maketitle

\begin{manuscriptsummary}
\sisetup{mode=text}\relax
Neodymium (Nd)-doped glasses, fibres and crystals~\cite{snitzer1961glass,koester1964amplification,geusic1964laser} emit near \SI{1}{\micro\meter} and enable applications ranging from ophthalmic surgery and nonlinear microscopy to laser machining and optical trapping of atoms for quantum science~\cite{ratkaytraub2003first,evans2005chemical,dubey2008laser,grimm2000optical,kaufman2021tweezer}.
Recent advances have realised such ion-doped solid-state sources in nearby spectral ranges using photonic integrated circuits, either by bonding crystals to chips or by using uniformly doped waveguide layers~\cite{wang2023photonic,yang2024titanium,guo2026ndmopa,zeng2026ytterbium}. 
Yet device complexity has remained limited: uniform doping places gain throughout the circuit, and crystal bonding is typically restricted to die-level processing. Both constraints hinder integrated tunable lasers and mode-locked femtosecond oscillators, which require gain sections combined with low-loss passive filters, mirrors and dispersion-engineered gratings.
Silicon nitride is a particularly attractive platform for such circuits: it is available from commercial foundries and space-compatible, and offers ultralow loss, high power handling and a wide transparency window~\cite{liuHighyieldWaferscaleFabrication2021,xiang2022silicon,brasch2014radiation}. However, net gain from rare-earth ions other than erbium has not been demonstrated in silicon nitride.
Here we overcome these challenges with selective ion implantation, which defines gain lithographically within a single waveguide layer and preserves unimplanted regions for passive circuitry, as developed for erbium-implanted silicon nitride~\cite{liuPhotonicIntegratedCircuit2022,liu2024fully,jilaser}. We demonstrate for the first time large gain coefficients of up to \SI{10}{dB/cm} in neodymium-implanted silicon nitride (Nd:\ce{Si3N4}) photonic integrated circuits, with high-power amplifiers, widely tunable continuous-wave Vernier lasers and femtosecond mode-locked oscillators. 
Amplifiers provide net gain across all three neodymium emission bands, 
with up to \SI{24.9}{dB} of on-chip gain, more than \SI{100}{mW} of output power and a photon conversion efficiency exceeding \SI{60}{\percent}.
Monolithic Vernier lasers are tunable over \SI{85.6}{nm}, with up to \SI{19.2}{mW} of on-chip output power and intrinsic linewidths below \SI{500}{Hz}.
Self-starting, passively mode-locked oscillators with a hybrid-integrated semiconductor saturable-absorber mirror operate at a \SI{2.37}{GHz} repetition rate, with an optical bandwidth supporting \SI{502}{fs} transform-limited pulses.
These results establish Nd:\ce{Si3N4} as a monolithic gain platform near \SI{1}{\micro\meter}, compatible with wafer-scale fabrication and offering a route to integration with nonlinear frequency conversion, electro-optic modulation and photodetection demonstrated in silicon nitride~\cite{churaev2023heterogeneous,cai2026heterogeneous,gherabli2023mose2}.

\end{manuscriptsummary}
\par\medskip

\begin{figure*}[p]
\centering
 	\includegraphics[width=0.95\textwidth,trim={0 3pt 0 0},clip]{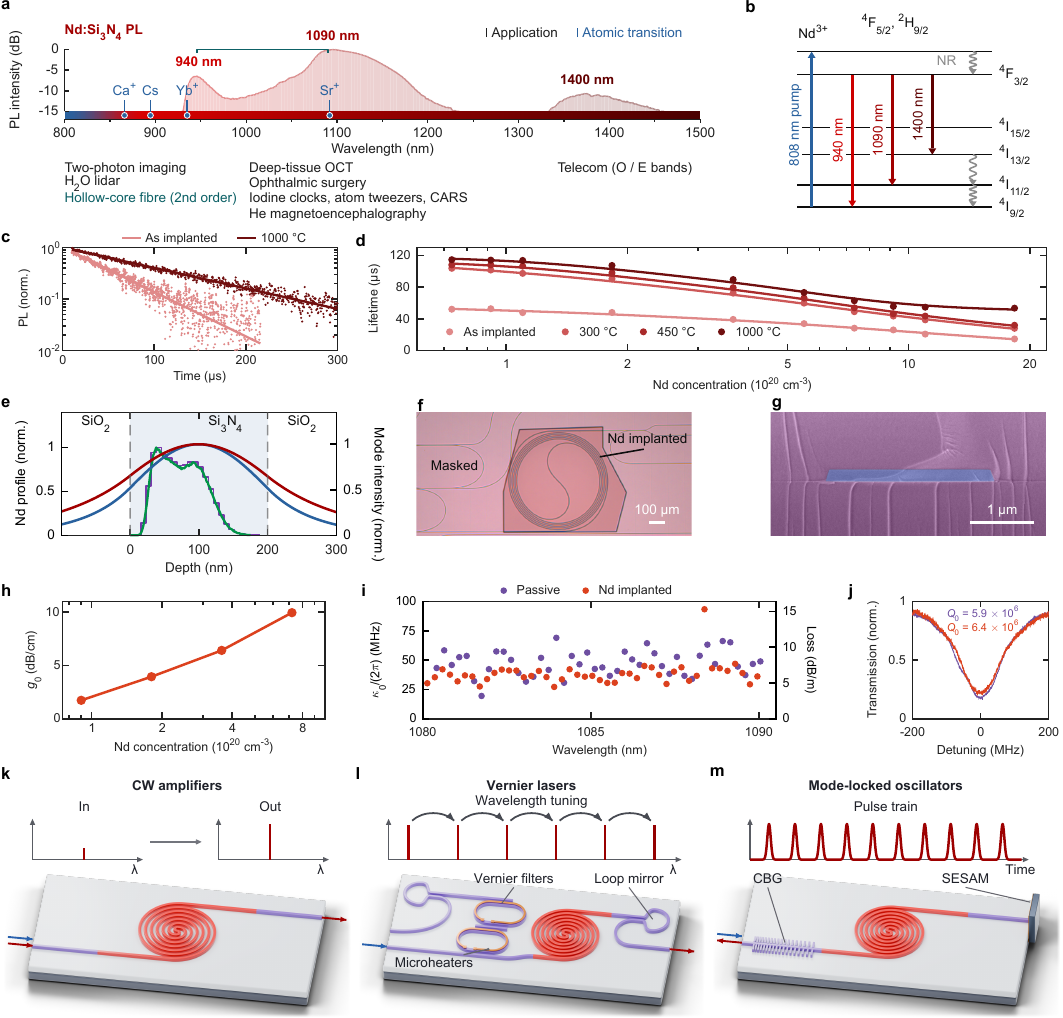}
    \setlength{\abovecaptionskip}{3pt}
 	\caption{\textbf{Neodymium ion-implanted photonic integrated circuits.}
    \textbf{a}, Nd:\ce{Si3N4} photoluminescence (PL) spectrum, near-infrared applications and atomic transitions (CARS, coherent anti-Stokes Raman scattering; OCT, optical coherence tomography)~\cite{zong2017miniature,wirth2009wales,unterhuber2005retinal,ratkaytraub2003first,roslund2024clocks,kaufman2021tweezer,evans2005chemical,fourcault2021helium,wang2021bismuth,bruzewicz2019trapped}.
    Teal bracket, second-order antiresonant window of hollow-core fibres, with losses below \SI{0.5}{dB/km}~\cite{chen2026hollowcore,sakr2021five}.
    \textbf{b}, Nd$^{3+}$ energy manifolds with the Nd:\ce{Si3N4} emission bands; NR, nonradiative relaxation.
    \textbf{c}, Photoluminescence decay of the as-implanted film and after annealing at \SI{1000}{\celsius}, with single-exponential fits.
    Peak Nd concentration: \SI{1.83e20}{cm^{-3}}.
    \textbf{d}, $^{4}\mathrm{F}_{3/2}$ lifetime versus peak Nd concentration as implanted and after annealing at \SI{300}{\celsius}, \SI{450}{\celsius} and \SI{1000}{\celsius}, showing concentration quenching.
    Curves guide the eye.
    \textbf{e}, Neodymium profile from Rutherford backscattering (purple), MONICA simulation~\cite{hellerMonica} (green) and fundamental transverse-electric (TE$_{00}$) mode intensities at \SI{808}{nm} (blue) and \SI{1090}{nm} (red).
    \textbf{f}, False-coloured micrograph of a selectively implanted spiral.
    \textbf{g}, False-coloured scanning electron micrograph of the Nd:\ce{Si3N4} waveguide cross-section.
    \textbf{h}, Peak internal gain coefficient $g_0$ versus peak Nd concentration.
    \textbf{i}, Intrinsic linewidths of passive (blue--purple) and implanted (red--orange) \SI{50}{GHz} microrings; the right axis gives propagation loss with group index $n_g=1.8$.
    Peak Nd concentration: \SI{1.80e20}{cm^{-3}}.
    \textbf{j}, Example resonances with similar intrinsic linewidths at \SI{1088.69}{nm} (passive) and \SI{1088.60}{nm} (implanted), with intrinsic quality factors $Q_0$ around 6 million.
    \textbf{k--m}, Schematics of the amplifier (k), Vernier laser (l) and mode-locked oscillator (m); red, Nd:\ce{Si3N4}; purple, passive \ce{Si3N4}.
    CBG, chirped Bragg grating; SESAM, semiconductor saturable-absorber mirror.
    Blue and red arrows denote pump and signal.}
 \label{Fig:1}

\end{figure*}

The near-infrared optical window around \SI{1}{\micro\meter} supports a broad range of applications (Fig.~\ref{Fig:1}a), including multiphoton microscopy, coherent anti-Stokes Raman scattering microscopy and optical coherence tomography in biological tissue~\cite{zong2017miniature,evans2005chemical,unterhuber2005retinal}. It is also used in water-vapour lidar~\cite{wirth2009wales} and atomic clocks, magnetometers and trapped-ion and neutral-atom quantum computers~\cite{roslund2024clocks,fourcault2021helium,bruzewicz2019trapped,kaufman2021tweezer}. 
In optical communication, growing data-centre traffic driven by artificial intelligence has increased interest in hollow-core fibres as candidates for future short-range optical interconnects~\cite{poletti2013towards,chen2025coherent}.
These fibres confine light to an air core through antiresonant reflection from thin glass membranes and have achieved losses of \SI{0.13}{dB/km} near \SI{1}{\micro\meter}~\cite{mahdiraji2026dual}, comparable to those of silica fibres in the C and L bands, thereby opening a new communication window that requires sources near \SI{1}{\micro\meter}.
Neodymium sources have provided coherent light in this region since the first demonstrations of neodymium-doped glass lasers, fibre amplifiers and neodymium-doped yttrium aluminium garnet (Nd:YAG) lasers~\cite{snitzer1961glass,koester1964amplification,geusic1964laser}.
Pumping near \SI{800}{nm} excites Nd$^{3+}$ ions from the $^{4}\mathrm{I}_{9/2}$ ground state to the $^{4}\mathrm{F}_{5/2}$ and $^{2}\mathrm{H}_{9/2}$ levels, which relax nonradiatively to the long-lived $^{4}\mathrm{F}_{3/2}$ manifold.
Transitions from $^{4}\mathrm{F}_{3/2}$ to the $^{4}\mathrm{I}_{9/2}$, $^{4}\mathrm{I}_{11/2}$ and $^{4}\mathrm{I}_{13/2}$ \mbox{manifolds} give three emission bands spanning roughly \SIrange{900}{1400}{nm}, with exact wavelengths that depend on the host (Fig.~\ref{Fig:1}b).

Photonic integrated circuits (PICs) with optical gain in this window have been explored as compact, high-performance devices that can be manufactured at wafer scale and bring amplifiers, lasers and femtosecond oscillators to applications in which size, weight, power and cost matter.
To date, however, on-chip gain and lasing near \SI{1}{\micro\meter} have largely relied on heterogeneously or hybrid-integrated III--V semiconductors~\cite{tran2022extending,Zhu2019DualBandHybrid,Chen2025AlGaAsSiN}.
Semiconductor gain responds on nanosecond or shorter timescales, transferring drive-current fluctuations to the optical output and causing excess noise.
By contrast, solid-state gain media such as rare-earth-doped crystals and fibres have upper-state lifetimes of microseconds to milliseconds. Their gain therefore acts as a low-pass filter for these fluctuations and introduces negligible interchannel crosstalk in amplifiers. Rare-earth-doped gain media also exhibit noise figures close to the quantum limit of \SI{3}{dB}, support high-power amplification and provide temperature-insensitive gain~\cite{liuPhotonicIntegratedCircuit2022,jilaser}.
Their weak gain--index coupling also avoids the linewidth enhancement characteristic of semiconductor lasers~\cite{henry1982theory}. 

In this spectral region, solid-state gain has been integrated using uniformly doped waveguide layers of silica~\cite{hibino1989neodymium}, tantalum pentoxide~\cite{unal2005ndta2o5}, aluminium oxide~\cite{yang2010neodymium,zeng2026ytterbium} and lithium niobate~\cite{he2026ndlnoi}, or Nd:YAG crystals bonded onto passive circuits~\cite{guo2026ndmopa}.
Passive components for wavelength-selective feedback and routing must therefore be fabricated either in the doped waveguides~\cite{bernhardi2010ultra} or in a separate waveguide layer coupled to the gain layer through interlayer transitions~\cite{Mu2020}.
In the doped layer, the geometry and spectral response of filters and mirrors are tied to the waveguide design of the gain section, whereas interlayer transitions add optical interfaces and fabrication steps.
Both configurations therefore constrain components that require precise spectral and dispersion design, such as Vernier microring filters for wavelength selection~\cite{liu2024fully,jilaser} and chirped Bragg gratings for dispersion-managed mode locking~\cite{li2026dispersionmanaged}.

Silicon nitride waveguides offer broad transparency and tailorable dispersion, with propagation losses of a few decibels per metre demonstrated at telecommunication wavelengths~\cite{liuHighyieldWaferscaleFabrication2021,Ji:24}.
They also provide a passive platform for heterogeneous integration of III--V gain media, lithium niobate and lithium tantalate~\cite{tran2022extending,churaev2023heterogeneous,cai2026heterogeneous}.
In the telecommunication band, selective ion implantation introduces erbium gain directly into silicon nitride.
A lithographic mask confines erbium to gain sections, leaving passive filters, mirrors and routing waveguides undoped in the same waveguide layer~\cite{liuPhotonicIntegratedCircuit2022,jilaser}.
This approach has enabled integrated amplifiers with output powers above \SI{100}{mW}~\cite{liuPhotonicIntegratedCircuit2022}, tunable lasers with intrinsic linewidths below \SI{100}{Hz}~\cite{liu2024fully,jilaser} and mode-locked oscillators with nanojoule pulse energies~\cite{qiu2025highpulseenergyintegratedmodelockedlasers} or full on-chip integration~\cite{li2026dispersionmanaged}.
The process has also been demonstrated at wafer scale using implantation energies below \SI{500}{keV}, compatible with industrial implanters~\cite{jilaser}.

However, whether this approach can be extended to other rare-earth ions is an open scientific question: the implanted ions must remain optically active and sustain net gain despite constraints that vary among ion species, including nonradiative decay mediated by phonons and implantation-induced defects, limited solubility, clustering at high concentrations and residual propagation loss~\cite{polmanErbiumImplantedThin1997,liuPhotonicIntegratedCircuit2022,sun2026cmos}.
Net gain from rare-earth ions in silicon nitride waveguides has so far been limited to erbium.
For neodymium, optical gain has been reported only at \SI{10}{K} in hydrogenated amorphous silicon-rich silicon nitride films~\cite{tessler2005gain}.
Ytterbium (Yb), which displaced neodymium in high-power fibre lasers owing to its small pump--signal quantum defect and freedom from excited-state absorption and concentration quenching~\cite{richardson2010high}, was estimated to have an optically active fraction of only a few per cent in one \ce{Si3N4} implantation study~\cite{dehaan2017spectroscopic}.

Here, we overcome these challenges and demonstrate the first neodymium-implanted \ce{Si3N4} (Nd:\ce{Si3N4}) PICs, combining selectively defined gain with unimplanted passive circuitry in one waveguide layer.
With optimised implantation conditions, the implanted waveguides reach internal gain coefficients of up to \SI{10}{dB/cm} and provide net gain in all three neodymium emission bands, with propagation losses comparable to those of passive \ce{Si3N4}.
On this platform, we realise three building blocks for integrated light sources: continuous-wave (CW) amplifiers with up to \SI{24.9}{dB} of on-chip gain and more than \SI{100}{mW} of output power, monolithic Vernier lasers tunable over \SI{85.6}{nm} with intrinsic linewidths below \SI{500}{Hz}, and self-starting, passively mode-locked oscillators operating at \SI{2.37}{GHz} with an optical bandwidth supporting femtosecond pulses.
These results establish a scalable neodymium-doped silicon nitride platform for amplifiers and lasers in the spectral region long served by bulk neodymium lasers.

\subsection*{Neodymium-doped silicon nitride photonic integrated circuits}

The photoluminescence (PL) spectrum of an Nd-implanted \ce{Si3N4} thin film (sample preparation in Supplementary Information, Section~1\,A) exhibits the three bands of the $^{4}\mathrm{F}_{3/2}$ manifold, with maxima around \SI{940}{nm}, \SI{1090}{nm} and \SI{1400}{nm} (Fig.~\ref{Fig:1}a,b).
All three are red-shifted relative to the corresponding bands of neodymium in oxide hosts, which lie near \SI{880}{nm}, \SI{1060}{nm} and \SI{1330}{nm} in aluminium oxide waveguides~\cite{yang2010neodymium} and at similar wavelengths in silica fibres~\cite{digonnet2013rareearth}.
This shift agrees with earlier observations of neodymium emission near \SI{1100}{nm} in silicon-rich silicon nitride~\cite{tessler2005gain,li2011ndsin} and is consistent with the more covalent bonding of the nitride host.
The bands are also broad: the $^{4}\mathrm{F}_{3/2}\rightarrow{}^{4}\mathrm{I}_{11/2}$ emission spans about \SI{70}{nm} at half maximum, compared with less than \SI{1}{nm} for the \SI{1064}{nm} line of Nd:YAG~\cite{koechner2006solidstate,guo2026ndmopa}.
In Nd:YAG all ions share one crystal field and thus one set of Stark levels, whereas in the amorphous nitride the Stark splitting varies from site to site and the components merge into inhomogeneously broadened bands, as in neodymium glasses~\cite{digonnet2013rareearth}.
The resulting broad gain spectrum, obtained at the price of a lower peak emission cross-section, is essential for the wide tuning and femtosecond mode locking reported below.
Annealing at \SI{1000}{\celsius} repairs the implantation damage and lengthens the $^{4}\mathrm{F}_{3/2}$ lifetime from \SI{48.0}{\micro\second} in the as-implanted film to \SI{107.3}{\micro\second} at a peak Nd concentration of \SI{1.83e20}{cm^{-3}}, calculated from the simulated implantation profile (Fig.~\ref{Fig:1}c; Supplementary Information, Sections~1\,B and~1\,C).
At higher Nd concentrations, obtained by increasing the implantation fluence, the lifetime shortens in every annealing state (Fig.~\ref{Fig:1}d): after annealing at \SI{1000}{\celsius} it falls from \SI{114.6}{\micro\second} at \SI{7.33e19}{cm^{-3}} to \SI{53.5}{\micro\second} at \SI{1.83e21}{cm^{-3}}.
Such concentration quenching is common in rare-earth-doped media: as the ion spacing decreases, excited ions lose their energy by cross-relaxation (F\"{o}rster energy transfer) with neighbouring ions and by energy migration to quenching sites~\cite{digonnet2013rareearth}.
It is also associated with excess optical loss~\cite{liuPhotonicIntegratedCircuit2022,qiu2025highpulseenergyintegratedmodelockedlasers}.

The Nd:\ce{Si3N4} PICs are defined by selective implantation of neodymium into low-loss silicon nitride waveguides, following the process developed for wafer-scale erbium-implanted \ce{Si3N4} (Er:\ce{Si3N4}) circuits~\cite{jilaser}.
Implantations at \SI{108}{keV}, \SI{193}{keV} and \SI{380}{keV} into \SI{200}{nm}-thick \ce{Si3N4} waveguides place the ions within the guided pump and signal fields (Fig.~\ref{Fig:1}e), at energies available on industrial implanters for full-wafer processing~\cite{jilaser}.
The calculated ion--mode overlap factors~\cite{liuPhotonicIntegratedCircuit2022}, using the implantation profile simulated with MONICA~\cite{hellerMonica}, are $\Gamma_{\mathrm{p}}=0.37$ at \SI{808}{nm} and $\Gamma_{\mathrm{s}}=0.31$ at \SI{1090}{nm} for a \SI{2.5}{\micro\metre}-wide waveguide.
A photoresist mask confines the implantation to the gain spirals, so that filters, mirrors and routing waveguides remain unimplanted in the same \ce{Si3N4} layer (Fig.~\ref{Fig:1}f,g; fabrication details in Supplementary Information, Section~1\,D). Their geometry and spectral response are therefore independent of the gain region, and no transition between waveguide layers is required.

Figure~\ref{Fig:1}h shows the peak internal gain coefficient $g_0$, i.e. the gain provided by the neodymium ions, as a function of peak Nd concentration.
It increases sublinearly with the Nd concentration and reaches up to \SI{10}{dB/cm} (Supplementary Information, Section~2\,D).
The losses of implanted and unimplanted waveguides are characterised by measuring the resonance linewidths of microring resonators (Fig.~\ref{Fig:1}i,j).
The fitted resonances give intrinsic linewidths $\kappa_0/2\pi$ near \SI{1090}{nm} of \SI{47.1 \pm 10.8}{MHz} for the passive microring and \SI{38.2 \pm 9.6}{MHz} for the implanted microring.
They correspond to propagation losses of \SI{7.7 \pm 1.8}{dB/m} and \SI{6.3 \pm 1.6}{dB/m}, respectively, using a group index $n_g=1.8$.
Uncertainties are standard deviations across the measured resonances.
The two values agree within their uncertainties, as expected for a four-level system in which the unpumped Nd ions do not absorb the signal, and show that the implantation and annealing leave no residual damage-induced loss.
The high gain coefficient (up to \SI{10}{dB/cm}) combined with low propagation losses (\SI{<10}{dB/m}) forms the basis for high-efficiency amplifiers and narrow-linewidth lasers~\cite{liuPhotonicIntegratedCircuit2022,liu2024fully}.

\begin{figure*}[htb!]
\centering
    \includegraphics[width=\textwidth]{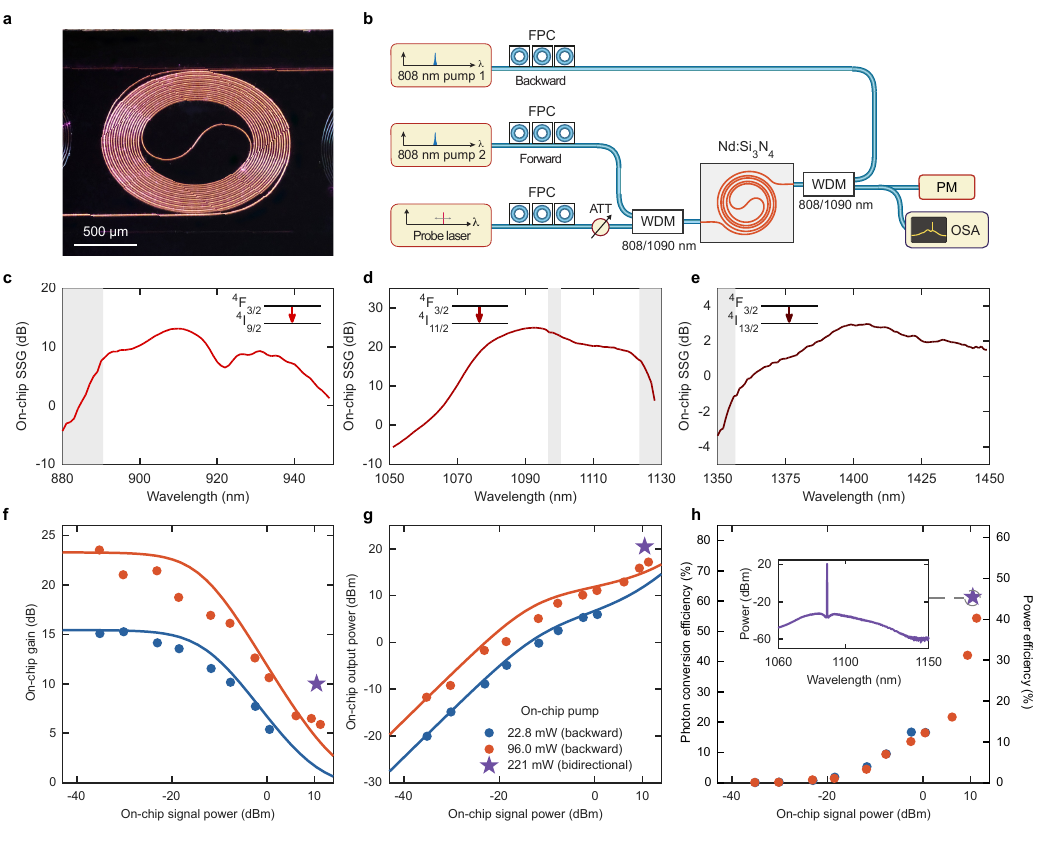}
    \caption{\textbf{Continuous-wave amplification on an Nd:\ce{Si3N4} waveguide amplifier.}
    \textbf{a}, Pumped Nd:\ce{Si3N4} \SI{7.7}{cm} spiral (scale bar, \SI{500}{\micro\metre}).
    \textbf{b}, Experimental setup for CW gain characterisation.
    FPC, fibre polarisation controller; ATT, attenuator; SLD, superluminescent diode; WDM, wavelength-division multiplexer; OSA, optical spectrum analyser; PM, power meter.
    \textbf{c--e}, TE-polarised on-chip small-signal gain (SSG) in the \SI{940}{nm}, \SI{1090}{nm} and \SI{1400}{nm} bands under \SI{96.0}{mW} of on-chip backward pump power.
    Insets show the Nd$^{3+}$ transitions; grey shading marks wavelengths where the amplified SLD signal is within \SI{1}{dB} of the amplified spontaneous emission (ASE).
    See Supplementary Information, Section~2 for measurement details.
    \textbf{f--h}, On-chip gain, output power and photon/power conversion efficiency versus input signal power.
    Solid curves in \textbf{f,g} show the CW amplifier model.
    The inset in \textbf{h} shows the calibrated on-chip pump-on spectrum for the circled bidirectional pumping point.
    Pump coupling loss is calibrated to be \SI{6.2}{dB} per facet, while signal coupling loss varies with wavelength from \SIrange{2.6}{5.6}{dB} per facet.
    Device ID: \texttt{D257\_D3\_C4}.}
    \label{Fig:3}

\end{figure*}

Figure~\ref{Fig:1}k--m shows the three device types demonstrated on this platform.
CW amplifiers use a single implanted spiral with a chosen doping concentration and length to deliver high gain and output power.
Placing the gain region in a linear cavity between two tunable loop mirrors, with a Vernier filter formed by two low-loss \ce{Si3N4} microring resonators and all elements tuned by microheaters, we demonstrate a monolithic widely tunable laser.
Passively mode-locked oscillators are realised through an integrated chirped Bragg grating (CBG), which provides broadband feedback and dispersion management, and a semiconductor saturable-absorber mirror (SESAM) hybrid-integrated at the other end of the cavity.

\subsection*{Continuous-wave amplification}

We characterise a \SI{7.7}{cm}-long Nd:\ce{Si3N4} spiral waveguide with a peak Nd concentration of \SI{1.80e20}{cm^{-3}}.
Figure~\ref{Fig:3}a shows a photograph of the pumped device, whose upconversion emission has a characteristic amber colour.
The amplifier is pumped at \SI{808}{nm}, with the pump and signal delivered through fibre wavelength-division multiplexers (WDMs) (Fig.~\ref{Fig:3}b).
Two probe sources are used: broadband superluminescent diodes (SLDs) measure the on-chip net small-signal gain (SSG), which includes on-chip propagation losses, and a narrow-linewidth tunable diode laser near \SI{1090}{nm} measures the gain and output power as functions of input signal power.

Figure~\ref{Fig:3}c--e shows the broadband SSG spectra for transverse-electric (TE) polarisation with \SI{96.0}{mW} of on-chip backward pump power.
The three bands correspond to emission from $^{4}\mathrm{F}_{3/2}$ to $^{4}\mathrm{I}_{9/2}$, $^{4}\mathrm{I}_{11/2}$ and $^{4}\mathrm{I}_{13/2}$, respectively.
The on-chip SSG reaches \SI{13.1}{dB} near \SI{910}{nm}, \SI{24.9}{dB} near \SI{1093}{nm} and \SI{3.0}{dB} near \SI{1405}{nm}.

With the tunable laser at \SI{1090.0}{nm}, increasing the input signal power saturates the amplifier, so the gain decreases as the output power rises (Fig.~\ref{Fig:3}f,g).
At \SI{96.0}{mW} backward pump power, the largest measured on-chip gain is \SI{23.5}{dB}, obtained at an input signal power of \SI{-35.2}{dBm}, consistent with the SSG measured with the broadband source.
With bidirectional pumping at \SI{221}{mW} of on-chip pump power, an on-chip input of \SI{10.48}{dBm} at \SI{1089.3}{nm} yields \SI{111.8}{mW} of output power, corresponding to an on-chip gain of \SI{10.0}{dB} (Fig.~\ref{Fig:3}f,g).
The output power is limited by the pump power delivered to the chip.
The power conversion efficiency from launched pump power to converted on-chip signal power is shown on the right axis of Fig.~\ref{Fig:3}h.
The photon conversion efficiency (PCE), which accounts for the quantum defect between the pump and signal wavelengths, is plotted on the left axis.
At their respective largest measured outputs, the PCE is \SI{54.4}{\percent} for the \SI{96.0}{mW} backward-pump condition and \SI{61.3}{\percent} for the bidirectional condition.
This efficiency is comparable with Er:\ce{Si3N4} amplifiers and other integrated rare-earth-doped amplifiers based on ytterbium or thulium (Supplementary Table~S3).
The small-signal noise figure is measured to be \SI{4.8}{dB} (Supplementary Information, Section~2\,C).
Calibration, simulation and measurement details are given in Supplementary Information, Section~2.

\begin{figure*}[htb!]
\centering
\includegraphics[width=\textwidth]{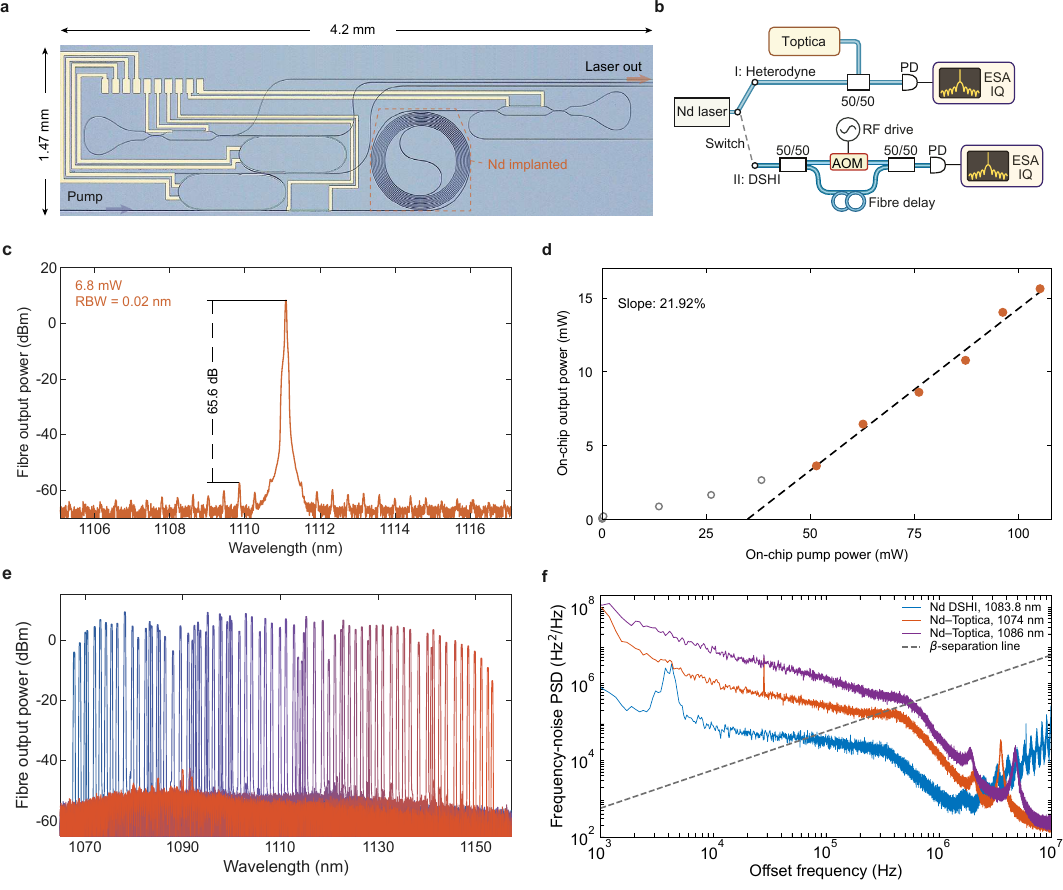}
\caption{\textbf{An Nd:\ce{Si3N4} photonic integrated widely tunable continuous-wave laser.}
\textbf{a}, Laser layout integrating an implanted gain spiral, two cascaded microring resonators forming a Vernier filter, and tunable loop mirrors.
Ti/Pt electrodes on the chip provide thermo-optic tuning of the microring resonances and of the mirror reflectivity.
\textbf{b}, Laser measurement setup for heterodyne detection and delayed self-heterodyne interferometry (DSHI).
PD, photodiode; AOM, acousto-optic modulator; RF, radio frequency; ESA, electrical spectrum analyser; IQ, in-phase and quadrature; PSD, power spectral density.
\textbf{c}, Representative single-mode fibre-output spectrum at \SI{1111.10}{nm} with calibrated \SI{6.8}{mW} output power in fibre.
The vertical bracket marks a \SI{65.6}{dB} contrast between the main peak and the strongest resolved secondary spectral feature.
The OSA resolution bandwidth (RBW) is \SI{0.02}{nm}.
\textbf{d}, Calibrated on-chip output power versus on-chip pump power for the state in \textbf{c}.
Filled dots denote the higher-power points used in the fit.
\textbf{e}, 76 independently optimised lasing states spanning \SIrange{1067.8}{1153.4}{nm}.
\textbf{f}, Frequency-noise spectra obtained by DSHI (Nd DSHI) at \SI{1083.8}{nm} and heterodyne detection against a Toptica reference laser (Nd-Toptica) at \SI{1074}{nm} and \SI{1086}{nm}.
The dashed grey line shows the $\beta$-separation line.
Device ID: \texttt{D257\_D3\_C1}.}
\label{Fig:4}
\end{figure*}
\subsection*{Tunable continuous-wave lasers}
Wavelength-selective feedback enables single-mode lasing near \SI{1090}{nm}.
The linear cavity in Fig.~\ref{Fig:4}a integrates an Nd:\ce{Si3N4} gain spiral, two tunable loop mirrors and a Vernier filter formed by two cascaded add--drop microring resonators in a single waveguide layer, following earlier Vernier laser designs~\cite{liu2024fully,jilaser}.
The lasers are pumped near \SI{808}{nm} and characterised with the setup in Fig.~\ref{Fig:4}b.
The cavity comprises a \SI{2.06}{cm} Nd:\ce{Si3N4} gain spiral, with the same peak Nd concentration (\SI{1.80e20}{cm^{-3}}) as the amplifier, and about \SI{0.4}{cm} of passive waveguide, giving a geometric length of \SI{2.46}{cm} and an estimated free spectral range (FSR) of \SI{3.4}{GHz}.
The two microrings have FSRs of \SI{100.441}{GHz}~($\Delta\nu_1$) and \SI{100.105}{GHz}~($\Delta\nu_2$), giving a Vernier interval of $\Delta\nu_{\mathrm{V}}\simeq \Delta\nu_1\Delta\nu_2/|\Delta\nu_1-\Delta\nu_2|\approx \SI{29.92}{THz}$, which covers the entire \SI{1090}{nm} emission band of Nd:\ce{Si3N4}.

Figure~\ref{Fig:4}c shows a single-mode spectrum at \SI{1111.10}{nm} with \SI{6.8}{mW} of output power in fibre, corresponding to an estimated on-chip power of \SI{15.6}{mW}. A weighted linear fit to the output power versus pump power yields an on-chip slope efficiency of \SI{22}{\percent} (Fig.~\ref{Fig:4}d). 
Thermo-optic tuning aligns the microring resonances at selected wavelengths across the neodymium gain spectrum.
Figure~\ref{Fig:4}e shows lasing states spanning \SIrange{1067.8}{1153.4}{nm} (\SI{85.6}{nm}).
The highest power in fibre is \SI{8.4}{mW} at \SI{1078.2}{nm}, which corresponds to an on-chip power of \SI{19.2}{mW}, exceeding the \SI{15}{mW} reported for an integrated Nd:YAG/\ce{Si3N4} master-oscillator power amplifier, without the downstream post-amplification~\cite{guo2026ndmopa}.
This span exceeds the reported tuning ranges of \SI{38}{nm} and \SI{63}{nm} of hybrid GaAs/\ce{Si3N4} and AlGaAs/\ce{Si3N4} Vernier lasers near \SI{1}{\micro\meter}, respectively~\cite{Zhu2019DualBandHybrid,Chen2025AlGaAsSiN}, and, in frequency (\SI{20.8}{THz}), is comparable to that of Ti:sapphire-on-insulator lasers~\cite{yang2024titanium} (Supplementary Table~S4).
Access to this range also depends on suppressing competing emission near \SI{1400}{nm}, as both transitions draw on the same upper state.
To favour lasing near \SI{1090}{nm}, the Vernier and loop-mirror biases are adjusted to increase the relative round-trip loss in the longer-wavelength band.
Adjusting these relative losses can instead favour emission near \SI{1400}{nm}, producing lasing in the telecommunication E band (Supplementary Information, Section~3) and indicating the potential for wavelength-selective lasing in either band through appropriate Vernier-filter and loop-mirror design.

Laser frequency noise is measured by heterodyne beating with a narrow-linewidth tunable reference (Toptica CTL 1050) and through delayed self-heterodyne interferometry (DSHI) (Fig.~\ref{Fig:4}b).
An electrical spectrum analyser (ESA) records the beat quadratures for phase and frequency reconstruction, followed by estimation of the single-sided frequency-noise power spectral density (PSD) using Welch's method~\cite{riebesehl2025interference}.
Figure~\ref{Fig:4}f compares heterodyne measurements at \SI{1074}{nm} and \SI{1086}{nm} with a DSHI measurement at \SI{1083.8}{nm}.
The DSHI spectrum reaches approximately \SI{1.3e3}{\hertz\squared\per\hertz} near \SI{1}{MHz}.
The heterodyne spectra reach minima of approximately \SI{1.3e2}{\hertz\squared\per\hertz} and \SI{1.4e2}{\hertz\squared\per\hertz} below \SI{10}{MHz}, corresponding to intrinsic Lorentzian linewidth estimates $\Delta\nu_{\mathrm{L}}=\pi S_{\nu,\min}$ of \SI{405}{Hz} and \SI{446}{Hz}, respectively.
At low offset frequencies, the delayed self-heterodyne spectrum exhibits lower noise, whereas the heterodyne spectra include an additional contribution from the reference laser.
At higher offset frequencies ($\gtrsim$ \SI{1}{MHz}), the high insertion loss of the DSHI setup reduces the radio-frequency (RF) beat-note signal-to-noise ratio and affects its sensitivity.
The two methods therefore provide complementary measurements across the offset-frequency range.
For an observation time of $T_0=\SI{1}{ms}$, full width at half maximum (FWHM) linewidths are estimated using the $\beta$-separation-line method~\cite{didomenico2010linewidth}, integrating the PSD above $S_\beta(f)=8\ln(2)f/\pi^2$ from $1/T_0=\SI{1}{kHz}$ to \SI{10}{MHz}.
The reconstructed DSHI spectrum at \SI{1083.8}{nm} yields \SI{194}{kHz}, and the heterodyne beats at \SI{1074}{nm} and \SI{1086}{nm} yield \SI{852}{kHz} and \SI{1.96}{MHz}, respectively. 

Supplementary Table~S4 presents the measured intrinsic linewidth of the Nd:\ce{Si3N4} laser alongside reported lasers at nearby wavelengths based on heterogeneous GaAs/\ce{Si3N4}~\cite{tran2022extending}, hybrid AlGaAs/\ce{Si3N4}~\cite{Chen2025AlGaAsSiN}, hybrid GaAs/\ce{Si3N4}~\cite{Zhu2019DualBandHybrid} and Ti:sapphire-on-insulator~\cite{yang2024titanium}.
Among these approaches, the Nd:\ce{Si3N4} laser achieves the lowest intrinsic frequency noise in a monolithic architecture.

\begin{figure*}[htb!]
\centering
    \includegraphics[width=\textwidth]{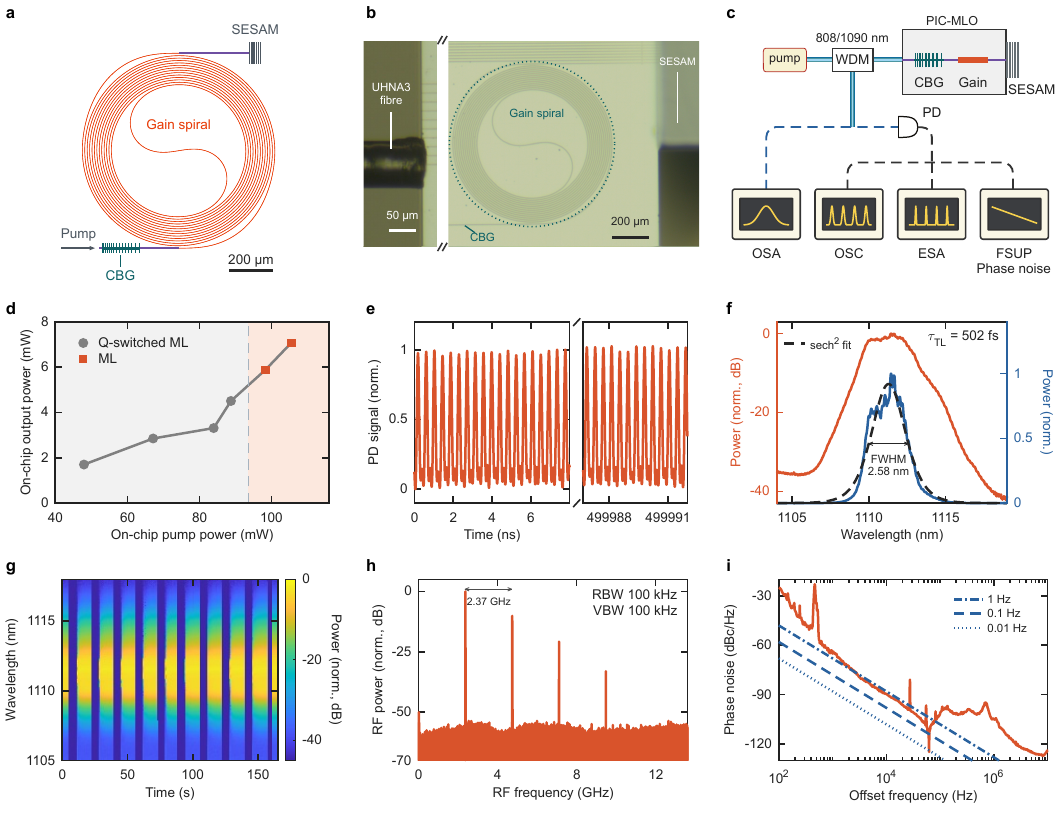}
    \caption{\textbf{An Nd:\ce{Si3N4} photonic integrated femtosecond oscillator.}
    \textbf{a}, Integrated mode-locked oscillator layout with Nd:\ce{Si3N4} gain spiral, chirped Bragg grating (CBG) and hybrid-integrated semiconductor saturable-absorber mirror (SESAM).
    \textbf{b}, Photograph of the device under measurement.
    \textbf{c}, Measurement setup with an \SI{808}{nm}/\SI{1090}{nm} wavelength-division multiplexer (WDM), photodiode (PD), optical spectrum analyser (OSA), oscilloscope (OSC), electrical spectrum analyser (ESA) and FSUP phase-noise analyser. MLO, mode-locked oscillator.
    \textbf{d}, On-chip output power versus on-chip pump power in the Q-switched and CW mode-locking regimes.
     \textbf{e}, Normalised photodiode signal from consecutive raw oscilloscope samples near the start and end of a \SI{500}{\micro\second} record, displayed using sinc interpolation.
    \textbf{f}, Normalised optical spectrum on logarithmic and linear scales; dashed line, $\mathrm{sech}^2$ fit indicating a transform-limited pulse duration of \SI{502}{fs}.
    \textbf{g}, Normalised optical spectra during repeated pump switching, showing self-starting operation.
    \textbf{h}, Normalised radio-frequency (RF) spectrum of repetition-rate harmonics.
    Resolution bandwidth (RBW) and video bandwidth (VBW) are both \SI{100}{kHz}.
    \textbf{i}, Single-sideband phase noise of the repetition rate, with reference lines for Lorentzian linewidths of \SI{1}{Hz}, \SI{0.1}{Hz} and \SI{0.01}{Hz}.
    Device ID: \texttt{D257\_D3\_C6}.}
    \label{Fig:5}

\end{figure*}

\subsection*{Femtosecond mode-locked oscillators}

The internal gain coefficient of Nd:\ce{Si3N4} (Fig.~\ref{Fig:1}h) allows a short gain section and therefore a compact cavity with a high repetition rate.
We use a peak Nd concentration of \SI{3.61e20}{cm^{-3}}, which gives a peak internal gain coefficient of \SI{6.4}{dB/cm}.
Following the dispersion-managed mode-locking architecture developed for Er:\ce{Si3N4}~\cite{li2026dispersionmanaged}, we form a linear cavity from an integrated CBG, an Nd:\ce{Si3N4} gain spiral of length \SI{2.4}{cm} and a SESAM~\cite{keller1996sesam} at the opposite chip facet (Fig.~\ref{Fig:5}a,b).
The CBG has \num{3000} periods, a centre waveguide width of \SI{1}{\micro\meter}, and a maximum one-sided width corrugation of \SI{300}{nm}.
Transfer-matrix simulations give the CBG a \SI{3}{dB} reflection bandwidth of about \SI{30}{nm} and a group delay dispersion (GDD) of \SI{-0.06}{ps^2}.
The waveguide section provides a round-trip GDD of \SI{0.04}{ps^2} and a round-trip internal gain of about \SI{30}{dB}.
The net round-trip GDD is about \SI{-0.02}{ps^2}, which is anomalous and supports dispersion-managed soliton mode locking~\cite{li2026dispersionmanaged}.
The \SI{808}{nm} pump is launched through the CBG, and the laser output, extracted from the same facet, is separated from the pump by an \SI{808}{nm}/\SI{1090}{nm} WDM (Fig.~\ref{Fig:5}c).

As the pump power increases, the oscillator passes from Q-switched mode locking to CW mode locking, which is established at an on-chip pump power of around \SI{100}{mW} (Fig.~\ref{Fig:5}d).
At an on-chip pump power of \SI{106}{mW}, the on-chip output power reaches \SI{7.1}{mW}, corresponding to an on-chip pulse energy of \SI{3.0}{pJ}.
The photodiode trace shows a stable pulse train throughout a \SI{500}{\micro\second} record (Fig.~\ref{Fig:5}e).
The optical spectrum is centred at \SI{1111.5}{nm}, and a $\mathrm{sech}^2$ fit yields a transform-limited pulse duration of \SI{502}{fs} (Fig.~\ref{Fig:5}f).
Mode locking is self-starting: when the pump is switched off and on ten times, the laser returns to the same mode-locked spectrum each time (Fig.~\ref{Fig:5}g).
The RF spectrum shows the fundamental repetition rate of \SI{2.37}{GHz} and its harmonics (Fig.~\ref{Fig:5}h), about twice that of Er:\ce{Si3N4} mode-locked oscillators based on the same architecture~\cite{li2026dispersionmanaged}.
Multi-gigahertz repetition rates are attractive for astronomical spectrograph calibration, where combs from conventional lasers must be filtered to line spacings of several to tens of gigahertz~\cite{steinmetz2008laser}.

The free-running repetition rate has a single-sideband phase noise of \SI{-89.5}{dBc/Hz} at \SI{10}{kHz} offset (Fig.~\ref{Fig:5}i), orders of magnitude lower than those reported for integrated semiconductor mode-locked oscillators at nearby wavelengths (Supplementary Table~S5).
Measurement details and long-term (\SI{40}{min}) stability of the integrated mode-locked oscillator are given in Supplementary Information, Section~4.

\subsection*{Discussion}
We have demonstrated Nd:\ce{Si3N4} PICs in which neodymium, implanted in lithographically defined regions, provides internal gain coefficients of up to \SI{10}{dB/cm} and net gain on all three $^{4}\mathrm{F}_{3/2}$ transitions in the same \ce{Si3N4} layer that forms the filters, mirrors and dispersion-managed cavities.
On this basis, we have realised the three building blocks of integrated light sources: amplifiers with more than \SI{100}{mW} of output power, Vernier lasers tunable over \SI{85.6}{nm} with intrinsic linewidths below \SI{500}{Hz}, and self-starting femtosecond oscillators at \SI{2.37}{GHz}.
Since the gain sections are defined by masked implantation after waveguide fabrication, all three device types share one process flow and are compatible with co-integration on a single chip and with wafer-scale production~\cite{jilaser}.
Uniformly doped films and bonded crystals do not readily offer this combination.
The long PL lifetime recovered after post-implantation annealing at only \SI{450}{\celsius} (Fig.~\ref{Fig:1}d) also paves the way to fully complementary metal-oxide-semiconductor (CMOS)-compatible fabrication on low-temperature silicon nitride platforms~\cite{zhang2024lowtemperature}.

Output power is currently limited by pump coupling losses above \SI{6}{dB}. Optimised inverse tapers with losses below \SI{1}{dB}~\cite{liu2018double} would raise the launched pump power, and with it the output, three- to four-fold. An on-chip amplifier after the tunable laser or mode-locked oscillator would boost it further.
Replacing the hybrid-integrated SESAM with an on-chip saturable absorber would make the mode-locked oscillators fully monolithic. Candidates include integrated graphene~\cite{reep2025graphene} and interferometric artificial saturable absorbers, which use the Kerr nonlinearity of the \ce{Si3N4} waveguide itself and need no additional material~\cite{shtyrkova2019integrated}.
With internal gain coefficients of up to \SI{10}{dB/cm}, a cavity about five times shorter is feasible and would raise the repetition rate above \SI{10}{GHz}, bringing rare-earth integrated mode-locked oscillators into the regime conventionally reached by soliton microcombs, used in optical communication and low-noise microwave generation~\cite{diddams2020optical}.

Hollow-core fibres with low-loss windows near both \SI{1}{\micro\meter} and \SI{1.55}{\micro\meter}~\cite{mahdiraji2026dual} could add a second transmission band to optical links. Amplifying both bands currently requires separate ytterbium- and erbium-doped fibre amplifiers~\cite{chen2026hollowcore}, whereas neodymium and erbium gain sections implanted in the same circuit could amplify both bands on one chip.
Silicon nitride also hosts heterogeneously integrated lithium niobate and lithium tantalate~\cite{churaev2023heterogeneous,cai2026heterogeneous} and efficient second-harmonic generation~\cite{wang2026integrated}, which could add fast electro-optic tuning, pulse picking and visible output to these sources.
Coherent anti-Stokes Raman scattering (CARS) microscopes typically pair a \SI{1064}{nm} neodymium laser with a synchronously pumped optical parametric oscillator~\cite{evans2005chemical}. On a chip, a neodymium mode-locked oscillator such as the one demonstrated here, paired with an erbium mode-locked oscillator~\cite{qiu2025highpulseenergyintegratedmodelockedlasers,li2026dispersionmanaged}, could instead directly provide the pump and Stokes pulses near \SI{1}{\micro\meter} and \SI{1550}{nm} for CARS microscopy of C--H vibrations in lipids and proteins~\cite{li2019multimodal}.

By combining the long-lived, low-noise gain of neodymium with the lithographic precision and wafer-scale manufacturing of low-loss silicon nitride, Nd:\ce{Si3N4} brings neodymium amplifiers, tunable lasers and mode-locked oscillators, so far realised mainly with bulk crystals and fibres, onto PICs.
This opens a route to compact light sources for biomedical imaging, atomic sensing and optical communication across the near infrared.

\pretolerance=0
\bigskip
\bibliographystyle{apsrev4-2}

	\footnotesize
\noindent\textbf{Funding Information:}
	This work was supported by the European Research Council (ERC) under the Horizon Europe research and innovation programme, grant agreement No. 101167540 (ATHENS), by the Deutsche Forschungsgemeinschaft (DFG) within project C5 of the CRC 1375 NOA (grant No. 398816777) and the large-scale equipment programme (grant No. 504643504), and by the Independent Research Fund Denmark (DFF), grant No. \mbox{5249-00033B}.

	\noindent\textbf{Acknowledgements:}
	The passive Si$_3$N$_4$ PIC fabrication and post-implantation processes were conducted in the EPFL Center of MicroNanoTechnology (CMi).
	We thank Di Che (Nokia Bell Labs) for useful discussions about optical communication at \SI{1}{\micro\meter}.

	\noindent\textbf{Author contributions:}
	J.S. and T.J.K. conceived the idea.
    J.S. and X.J. performed experiments, simulations and analysis related to \ce{Si3N4} chips.
    J.S. designed the \ce{Si3N4} chips.
    X.J., M.A., and Y.Z. fabricated the \ce{Si3N4} chips.
    S.L., A.K., and L.K. performed the ion implantation and related simulation and analysis with the help of J.S.
    Z.Q., Y.L., X.Y., Y.X., and X.L. provided experimental support.
    J.S. and X.J. wrote the manuscript with contributions from all coauthors.
    T.J.K., C.R. and Z.Q. supervised the work.

	\noindent\textbf{Competing interests:}
	T.J.K. is a cofounder and shareholder of LiGenTec SA, a start-up company offering Si$_3$N$_4$ photonic integrated circuits as a foundry service.
	T.J.K. is a cofounder and shareholder of EDWATEC SA, a start-up company offering optical amplifiers on chip.

	\noindent\textbf{Materials \& Correspondence:}
	Correspondence and requests for materials should be addressed to T.J.K.

	\noindent\textbf{Data and code availability:}
	The code and data used in this work will be provided in a \texttt{Zenodo} repository upon publication of this manuscript.

\normalsize
\bibliography{ref}
\end{document}